\documentclass[aps,prl,twocolumn]{revtex4}
\usepackage{graphics,graphicx,amssymb,amsmath,epstopdf}
\begin{document}

\title{Probing electronic order via coupling to low energy phonons in superconducting Bi$_2$Sr$_{2-x}$La$_{x}$CuO$_{6+\delta}$}
\author{C.J.~Bonnoit$^{1}$}
\author{D.R.~Gardner$^{1}$}
\author{R.~Chisnell$^{1}$}
\author{A.H.~Said$^{2}$}
\author{Y.~Okada$^{1,3}$}
\author{T.~Kondo$^{1,4}$}
\author{T.~Takeuchi$^{3}$}
\author{H.~Ikuta$^{3}$}
\author{D.E.~Moncton$^{1}$}
\author{Y.S.~Lee$^{1*}$}
\affiliation{$^{1}$Department of Physics, Massachusetts Institute of Technology, Cambridge, MA 02139}
\affiliation{$^{2}$Advanced Photon Source, Argonne National Laboratory, Chicago, IL 60439}
\affiliation{$^{3}$Department of Crystalline Materials Science, Nagoya University, Nagoya 464-8603, Japan}
\affiliation{$^{4}$Institute for Solid State Physics (ISSP), University of Tokyo, Kashiwa, Chiba 277-8581, Japan}
\date{\today}

\begin{abstract}
We report high-resolution inelastic x-ray scattering measurements of the acoustic phonons in the single-layer cuprate Bi$_2$Sr$_{2-x}$La$_{x}$CuO$_{6+\delta}$.  These measurements reveal anomalous broadening of the longitudinal acoustic phonon near the $(\frac{1}{4},\frac{1}{4},0)$ wavevector.  The observed wavevector and its doping dependence indicate the coupling of the phonons to an underlying electronic density wave state.  In addition, a comparison of the scattered intensities for x-ray energy-gain and x-ray energy-loss suggests that both time-reversal and inversion symmetries are broken in the material.  Upon cooling, the effects of symmetry breaking are enhanced in the pseudogap state.
\end{abstract}

\maketitle

In the cuprate superconductors there is an intriguing interplay between different types of electronic order which may compete and/or coexist with the superconductivity.  The phase diagram displays an unusual pseudogap region which exists below a temperature $T^*$ and is likely characterized by a symmetry-breaking electronic state.   A rich variety of broken-symmetries have been observed in different materials, including translation \cite{Tranquada,Vojta}, time-reversal \cite{Bourges,Kapitulnik,Greven}, four-fold rotation \cite{Hinkov,Taillefer,Lawler}, and inversion \cite{Kubota}. Understanding how these broken-symmetries relate to each other and which ones are universal is one of the key questions under current debate.

Examples of order that break translational symmetry are the stripe-like states of spin and charge seen with neutron and x-ray scattering \cite{Tranquada,Tranquada_xray} and the checkerboard pattern of the local density of states seen with scanning tunneling microscopy (STM) \cite{Hoffman,Howald,Vershinin,Hudson}. Typically, an electronic state which breaks translational symmetry should result in a static distortion of the crystal lattice if the electron-phonon coupling is strong enough. However, such electronic density wave order has not been observed in the same material by both STM and bulk scattering probes.  Possible explanations point to the surface sensitivity of the STM measurements or a short correlation length for the lattice distortion.\cite{Smadici}  Our results clarify this issue by showing that the lattice modulation remains dynamic down to low temperatures.  Furthermore, the scattered intensity from the phonons indicates that the combination of inversion (I) and time-reversal (TR) symmetries are broken at low temperatures.

Single crystals of the high-$T_c$ superconductor Bi$_2$Sr$_{2-x}$La$_{x}$CuO$_{6+\delta}$ (Bi2201) were examined with inelastic x-ray scattering with a high energy-resolution of $\sim 1.5$ meV to measure the phonon excitations, using the HERIX instrument at Sector 30 of the Advanced Photon Source at Argonne National Laboratory.  Three different compositions were studied: two samples with $T_c=25~\hbox{K}$ and $T_c=31~\hbox{K}$, referred to as samples La25K and La31K, respectively, and a third, Pb-doped sample with $T_c=33~\hbox{K}$, referred to as LaPb33K.  Details of sample preparation are described elsewhere \cite{Okada1,Okada2, Kondo}.  The Pb-doping is used to suppress the well-known structural incommensurate supermodulation with wavevector \mbox{(0,$\sim$0.23,1)} in orthorhombic notation. All three samples are found to consist of a single structural domain. Note that the in-plane direction of the structural supermodulation is at $45^\circ$ to the wavevector of the STM checkerboard order which is near $(\frac{1}{4},\frac{1}{4},0)$ in orthorhombic notation.

\begin{figure}[h!]
\includegraphics[width=8.5cm]{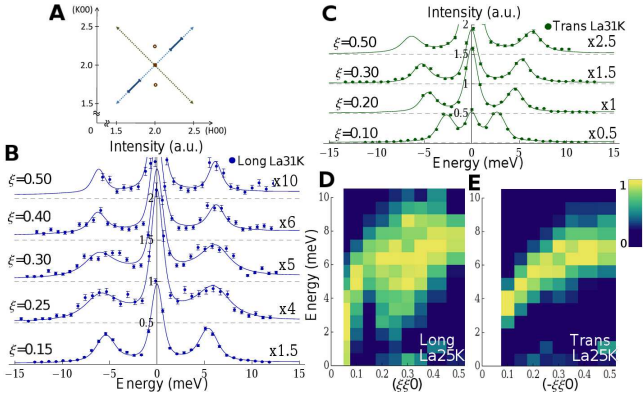} \vspace{0mm}
\caption{\label{Figure1} Acoustic phonon measurements.  (A) The dashed lines indicate the cuts near the $(2,2,0)$ Bragg position along which the phonons were measured. The longitudinal scans correspond to $(2,2,0) \pm (\xi,\xi,0)$ and the transverse scans correspond to $(2,2,0) \pm (-\xi,\xi,0)$. The solid bars correspond to the positions of the phonon anomaly and the open circles correspond to the in-plane position of the incommensurate superlattice peaks at $(2,2 \pm 0.23,1)$.  (B) Energy scans of the longitudinal acoustic phonon for the La31K sample.  (C) Energy scans of the transverse acoustic phonon for the same sample. The lines denote fits to a damped harmonic oscillator response function as described in the text.  (D) Longitudinal acoustic phonon data for the La25K sample, and (E) transverse acoustic phonon data, plotted as an intensity contour map where the background-subtracted intensity is normalized to the peak intensity for each energy scans.}
\end{figure}

Figure 1(A) depicts the trajectories in reciprocal space along which the phonons were measured. The polarization dependence of the inelastic x-ray scattering cross section from phonons, which is proportional to $(\vec{Q} \cdot \vec{\epsilon})^2$, allows one to separately measure the longitudinally and transversely polarized modes. For the La31K sample, energy scans of the longitudinally polarized phonons are shown in Fig.~1(B), and scans of the transversely polarized phonons are in Fig.~1(C).  A principal result of this paper is that an anomalous broadening of the longitudinal mode is observed for reduced wavevector $\vec{q} \simeq (0.25,0,25,0)$, which is not observed in the transverse mode.   This behavior is consistent with coupling of the lattice to an electronic density wave with characteristic wavevector near $(\frac{1}{4},\frac{1}{4},0)$ as suggested by previous STM \cite{Hudson} and ARPES studies \cite{Hashimoto}.  Measurements on the La25K sample reproduce this feat ure, as shown in Fig.~1(D) and 1(E).  All the data depicted in Figure 1 were taken at $T=300$~K.

This acoustic phonon anomaly is qualitatively different than the previously reported optical phonon anomalies in the high-$T_C$ cuprates \cite{Vojta}. A broadening of optical modes is observed in several cuprates near the same $\sim(\frac{1}{4},\frac{1}{4},0)$ reduced wavevector that we observe for the acoustic phonon.  While a period-four stripe state has been often invoked to explain this feature, many of the materials which display this optical phonon anomaly are not known to contain static stripe order.  The energy of the acoustic mode that we measure is about an order of magnitude smaller than that of the optical mode anomaly previously observed in Bi2201 \cite{Graf}.  And correspondingly, the relative broadening of the acoustic mode is significantly stronger (by about a factor of 3).

\begin{figure}[h!]
\includegraphics[width=8.5cm]{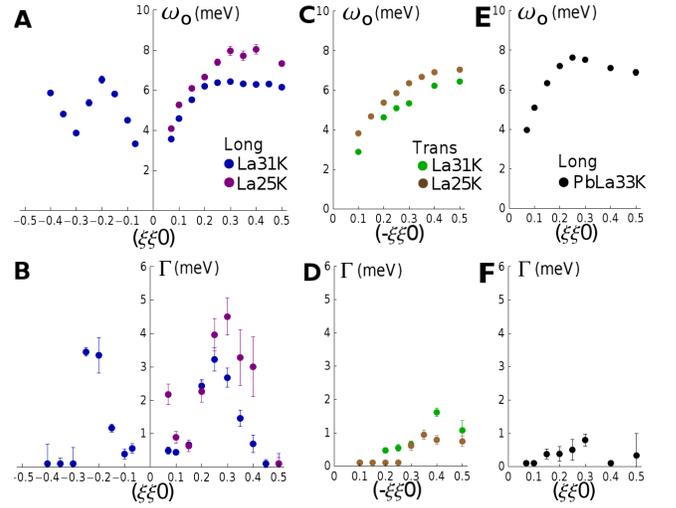} \vspace{0mm}
\caption{\label{Figure2} The phonon frequency $\omega_\circ$ and linewidth $\Gamma$ which results from fits as described in the text. (A)\&(B) the longitudinal modes in La31K and La25K, (C)\&(D) the transverse modes in La31K and La25K, (E)\&(F) the longitudinal mode in PbLa33K.  The anomalously large linewidth for the longitudinal modes in panel (B) is not seen in the transverse modes of the same samples nor in the longitudinal mode of the Pb-doped sample.}
\end{figure}

The scattered intensity from the acoustic phonon is proportional to the dynamic structure factor $S(\vec{Q},\omega)$, which was fit to a damped harmonic oscillator response function convoluted with the instrumental resolution. The energy resolution function was determined by fitting the elastic signal to a pseudo-Voight lineshape.  A two-dimensional convolution was performed over the momentum component along the scan direction and the energy.  A small background was included, which involves a constant term plus another term arising from inelastic scattering processes from the sample which is flat in energy but constrained to follow detailed balance.  We extracted values for the undamped phonon frequency $\omega_\circ$ and the damping parameter $\Gamma$ (proportional to the phonon's linewidth in energy), and the results are shown in Fig.~2(A) and (B) for the longitudinal mode, and in Fig.~2(C) and (D) for the transverse mode.  There is a clear anomaly in the linewidth of the longitudinal mode of both samples which is peaked at a wavevector around (0.25, 0.25, 0).  This anomaly does not occur in the transverse mode, as previously noted.  Since the longitudinal polarization corresponds to deformations of the bond lengths while the transverse polarization does not strongly change bond lengths, the former should be much more strongly coupled to any density wave state.

Interestingly, the anomaly is absent in the sample doped with Pb, as shown in Fig. 2(E) and 2(F).  The reason why the phonon anomaly is observed only in samples without Pb-doping is not completely clear.  A possible explanation is that the presence of the structural supermodulation enhances the electron-phonon coupling.  Raman measurements of high energy optical phonons in the related Bi2212 superconductor indicate that Pb-doping narrows the optical phonon, which is interpreted as resulting from a reduced electron-phonon coupling due to suppression of the supermodulation \cite{Opel}. In fact, an increase in electron-phonon interaction due to the presence of an incommensurate modulation have been reported in a prototypical CDW system \cite{Clerc}.  In our study, the dependence on Pb-doping suggests that the coupling of the lattice to the electronic order is particularly sensitive to the details of the crystal structure.  Note that the STM measurements see the underlying electronic density wave irrespective of Pb-doping \cite{Hudson,Kurosawa}.

In Fig.~3, the fitted intensity of the phonon at zero energy transfer, $S(\vec{Q},\omega=0)$, is shown.  Direct measurement of the elastic scattering does not reveal a clear peak as a function of $\vec{Q}$ due to the strong elastic background which dominates the signal.  Here, by measuring the phonon spectra away from $\omega=0$, the extension of spectral weight from the strongly damped phonon to zero energy can be deduced.  This intensity is quite different than having a sharp peak centered at $\omega=0$, such as a Bragg peak or a central peak. Based on the widths of the peaks, we estimate that the ``correlation length'' of the phonon anomaly is roughly 20 {\AA} within a finite-size domain model.  Also shown in Fig.~3 is the doping-dependent shift seen in the Fourier-transformed STM amplitude which measures the local density of states\cite{Hudson}.  While the magnitude of the wavevector seen with x-rays is slightly larger than that seen with STM, the doping dependent shifts  are nearly identical.   We note that additional longitudinal phonon modes along the $b^*$ direction have been observed in Bi2201 which arise from the incommensurate superstructure.\cite{Keimer}  The observed doping dependence of the anomaly, as well as the fact that only the longitudinal mode is broadened, argues that such additional modes do not explain the anomaly that we observe.

\begin{figure}[t!]
\includegraphics[width=8.2cm]{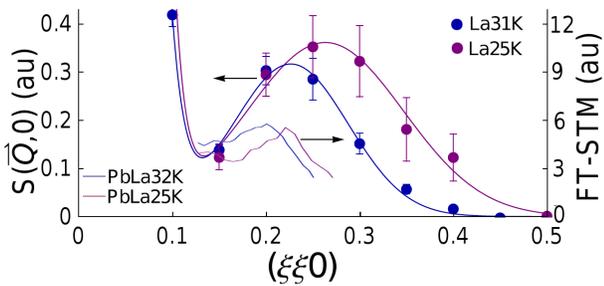} \vspace{0mm}
\caption{\label{Figure3} Sample dependence of the characteristic wavevector of the phonon anomaly. $S(\vec{Q},\omega=0)$, the scattered intensity at zero energy, is compared with the Fourier-transformed STM amplitudes for samples with comparable doping levels \cite{Hudson}.}\vspace{0mm}
\end{figure}

The phonon data presented so far were taken above $T^*$ for both samples ($T^*\approx 250$~K for La25K and $T^*\approx 150$~K for La31K \cite{Okada2}), and presumably reflect the effects of coupling to fluctuations of the electronic density wave.  Such coupling to dynamic degrees of freedom is reminiscent of the soft phonon behavior in the spin-Peierls compound TiOCl, where the phonons are strongly broadened well above the transition temperature \cite{Abel}.  However, in contrast to a conventional soft phonon transition, the frequency and linewidth of the longitudinal phonons in Bi2201 do not significantly change upon cooling to $T=100$~K, as shown in Fig.~4(A,B).  Also, neither a Bragg peak nor a central peak appears upon cooling to $T=5$~K, indicating the absence of any static lattice distortion at the anomaly wavevector.

\begin{figure}[t!]
\includegraphics[width=7.4cm]{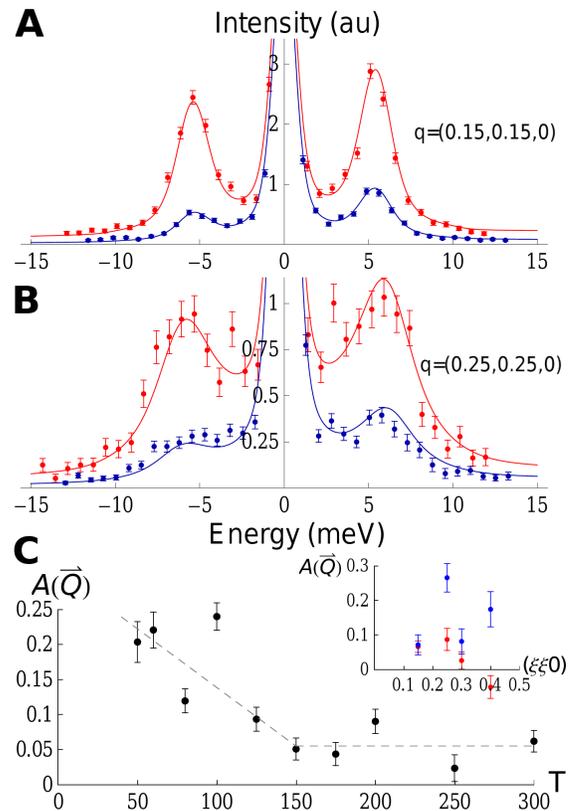} \vspace{0mm}
\caption{\label{Figure4} The intensity deviation from the conventional expectations of detailed balance.  Energy scans of the longitudinal acoustic phonons in La31K at $\vec{Q}=(2,2,0)+(\xi,\xi,0)$ at temperatures $T=300$~K and $T=100$~K for (A) $\xi=0.15$ and (B) $\xi=0.25$.  The solid lines denote fits which reflect detailed balance in the presence of inversion symmetry or time-reversal symmetry.  For the $\xi=0.25$ phonon mode at $T=100$~K, the asymmetric deviation from the line indicates that both I and TR symmetries are broken.  (C) The asymmetry ratio $A(\vec{Q})$ at $\vec{Q}=(2.25,2.25,0)$ as a function of temperature.  Inset: $A(\vec{Q})$ at $T=300$~K (red) and 100~K (blue).  The data show a clear enhancement of the asymmetry upon cooling.}\vspace{0mm}
\end{figure}

There is, however, an intriguing effect observed in the intensity of the phonons at low temperatures.  At the wavevector $\vec{Q}=(2.25,2.25,0)$ at $T=100$~K, there is a deviation of the intensity from conventional expectations.  For a system in thermal equilibrium, the fundamental principle of detailed balance requires that $S(-\vec{Q},-\omega)=\exp(-\frac{\hbar \omega}{k_B T}) \; S(\vec{Q},\omega)$.  If the system obeys the relation $S(-\vec{Q},\omega)=S(\vec{Q},\omega)$, then the intensity balance holds for an energy scan at fixed $\vec{Q}$.  This relation can only be violated if \emph{both} inversion (I) and time-reversal (TR) symmetries are broken in the system \cite{Lovesey,Kuscer}.  The solid lines in Fig.~4(A,B) denote fits in which either I or TR symmetry is preserved.  There is a statistically significant asymmetry in the intensity for the low temperature data at $\vec{Q}=(2.25,2.25,0)$, where the negative energy transfer data fall above the line (and vice versa for positive energy transfers).  This effect is not simply related to the crystal structure having a non-centrosymmetric unit cell since TR must also be broken (additionally, the literature reports both centrosymmetric and non-centrosymmetric unit cells for Bi2201 \cite{Ito,Gladyshevskii}).  This asymmetry of the intensity has a clear temperature dependence between $T=300$~K and $T=100$~K (which spans $T^* \approx 150$~K), showing that the broken symmetries are enhanced in the pseudogap phase.  In addition, the symmetries are not broken uniformly for all wavevectors, as the phonon at $\vec{Q}=(2.15,2.15,0)$ does not show the same magnitude of asymmetry.  This suggests that the broken I and TR symmetries are related to the electronic density wave state.  For example, a density wave state which is neither site-centered nor bond-centered would break inversion symmetry.  Additional orderings which would also break TR symmetry include circulating currents \cite{Varma} and incommensurate spin density waves \cite{Kimura}.

In order to better quantify the asymmetry, we examine the following asymmetry ratio $A(\vec{Q})=\frac{\chi^{\prime\prime}_+(\vec{Q}) - \chi^{\prime\prime}_-(\vec{Q})}{\chi^{\prime\prime}_+(\vec{Q}) + \chi^{\prime\prime}_-(\vec{Q})}$, where $\chi^{\prime\prime}_\pm(\vec{Q})$ denotes the magnitude of the imaginary susceptibility for the longitudinal acoustic phonon for $\pm$ energy transfers, computed using the following procedure. The raw data for each energy scan is background-corrected by subtracting a fitted pseudo-Voight elastic component and a small background component, as discussed previously.  The same background parameters were used at all temperatures.  After dividing by the appropriate thermal factor, the data represent the imaginary susceptibility via the fluctuation-dissipation theorem.  An energy integration was performed by summing the data from $-3~\hbox{meV}$ to $-10~\hbox{meV}$ to obtain $\chi^{\prime\prime}_-(\vec{Q})$ and from $3~\hbox{meV}$ to $10~\hbox{meV}$ to obtain $\chi^{\prime\prime}_+(\vec{Q})$.  Here, we represent $\chi^{\prime\prime}_\pm(\vec{Q})$ by their magnitudes.  We note that modest changes to the integration range (within a couple meV) do not significantly affect the magnitude of $A(\vec{Q})$, nor do minor changes to the background parameters.

Assuming that the principle of detailed balance holds (which is the case if the scattering lengths are real, as in the Thomson x-ray cross-section), then a nonzero value of $A(\vec{Q})$ can occur if and only if both I and TR symmetries are broken.   As shown in Fig.~4(C), this ratio clearly deviates from zero at low temperature for $\vec{Q}=(2.25,2.25,0)$. In fact, $A(\vec{Q})$ appears to be nonzero even at $T=300$~K, and it significantly increases upon cooling below $T^* \simeq 150$~K into the pseudogap state. As shown in the inset in Fig.~4(C), the asymmetry is most clearly apparent at the reduced wavevector $(0.25,0.25,0)$.

The combination of our observations with other measurements in the literature reveals an intriguing picture for the electronic state below $T^*$ in the pseudogap phase. Recent studies of Bi2201 using ARPES and optical techniques show that a phase transition occurs below $T^* \simeq 130$~K to a symmetry-breaking electronic state \cite{He}.  Our results indicate that I and TR symmetries are at least two of the symmetries broken at low temperatures.  The lattice modulation remains slowly fluctuating above and below $T^*$, so it is not clear from our measurements whether translational symmetry is spontaneously broken at $T^*$.  The conclusion of broken TR symmetry is consistent with previous Kerr rotation observations in Bi2201\cite{He}, and recent calculations suggest that I symmetry may be broken as well\cite{Orenstein}. In our x-ray results, the broken symmetries are most clearly observed at the wavevector of the anomalous phonon, although it seems likely that inversion symmetry is broken for the average unit cell as well.  While the lattice is probably only weakly coupled to the underlying electronic state, measurements of the low energy phonons, nonetheless, provide an unexpectedly valuable probe of the symmetries within the pseudogap phase.

We acknowledge E.W. Hudson, S. Todadri, P.A. Lee, and J.E. Hoffman for valuable discussions, and T.H. Han for assistance in the experiments.  The work at MIT was supported by the Department of Energy (DOE-BES) under Grant No. DE-FG02-07ER46134. Use of the Advanced Photon Source at Argonne National Laboratory was supported by the DOE-BES under Contract No. DE-AC02-06CH11357. The construction of HERIX was partially supported by the NSF under Grant No. DMR-0115852.

\end{document}